\PassOptionsToPackage{english}{babel}  
\documentclass[%
 reprint,
 amsmath,amssymb,
 aps,
]{revtex4-2}
\usepackage{babel}  
\usepackage{blindtext}
\usepackage{graphicx}
\usepackage{dcolumn}
\usepackage{bm}

\begin{document}

\preprint{APS/123-QED}

\title{Momentum flatband and superluminal propagation in a photonic time Moiré superlattice}

\author{Linyang Zou\textsuperscript{1}}
\author{Hao Hu\textsuperscript{2}}
\author{Haotian Wu\textsuperscript{1}}
\author{Yang Long\textsuperscript{3}}
\author{Yidong Chong\textsuperscript{3,4}}
\author{Baile Zhang\textsuperscript{3,4}}
\email{blzhang@ntu.edu.sg}
\author{Yu Luo\textsuperscript{1,2}}
\email{yu.luo@nuaa.edu.cn}
 
\affiliation{
\textsuperscript{1}School of Electrical and Electronic Engineering, Nanyang Technological University, Singapore 639798, Singapore.\\\textsuperscript{2}Key Laboratory of Radar Imaging and Microwave Photonics, Ministry of Education, College of Electronic and Information Engineering, Nanjing University of Aeronautics and Astronautics, Nanjing 211106, China.\\\textsuperscript{3} Division of Physics and Applied Physics, School of Physical and Mathematical Sciences, Nanyang Technological University, Singapore 637371, Singapore.\\\textsuperscript{4} Centre for Disruptive Photonic Technologies,
Nanyang Technological University, Singapore, 637371, Singapore.
 }






\begin{abstract}
Flat bands typically describe energy bands whose energy dispersion is entirely or almost entirely degenerate. One effective method to form flat bands is by constructing Moiré superlattices. Recently, there has been a shift in perspective regarding the roles of space (momentum) and time (energy) in a lattice, with the concept of photonic time crystals that has sparked discussions on momentum dispersion such as the presence of a bandgap in momentum. Here we propose a photonic time moiré superlattice achieved by overlaying two photonic time crystals with different periods. The resulting momentum bandgap of this superlattice supports isolated momentum bands that are nearly independent of energy, which we refer to as momentum flat bands. Unlike energy flat bands, which have zero group velocity, momentum flat bands exhibit infinitely large group velocity across a broad frequency range. Unlike previous optical media supporting broadband superluminal propagation based on gain, the effective refractive index of the momentum flat bands is real-valued, leading to more stabilized superluminal pulse propagation.  
\end{abstract}

\maketitle


\section{INTRODUCTION}
Photonic time crystals (PTCs), a temporal analogue of spatial photonic crystal (SPCs), have attracted extensive attention in the area of optics and photonics recently. Owing from the space-time duality in maxwell equation, modulating the media in time provides another degree of freedom for designing space-time metamaterials. As an analog of spatial boundary, it was shown theoretically \cite{Morgenthaler_1958,Akbarzadeh_Chamanara_Caloz_2018,Vezzoli_Bruno_DeVault_Roger_Shalaev_Boltasseva_Ferrera_Clerici_Dubietis_Faccio_2018,Pacheco-Peña_Engheta_2020b} and experimentally \cite{Moussa_Xu_Yin_Galiffi_Ra’di_Alù_2023,Zhou_Alam_Karimi_Upham_Reshef_Liu_Willner_Boyd_2020} that a temporal interface, where the wave impedance of the system varies in time while remains uniform in space, can induce reflection and refraction of electromagnetic waves. In the temporal scattering scenario, all the scattered waves emerge at the same side of interface and share the same parity as incident wave, which is fundamentally different with a conventional spatial scattering. Furthermore, in PTCs where the modulation is periodic in time, the interference of successional reflected and refracted wave can give rise to Floquet modes and form temporal photonic band structure \cite{Zurita-Sánchez_Halevi_Cervantes-González_2009}, opening bandgaps in momentum space (known as momentum bandgaps) \cite{Lustig_Sharabi_Segev_2018}, instead of in energy space (known as energy bandgaps) in spatial photonic crystals \cite{Fink_Winn_Fan_Chen_Michel_Joannopoulos_Thomas_1998,Yablonovitch_1993},. Unlike the decaying bandgap mode in conventional energy gap, in the momentum bandgaps, temporal modulation breaks the continuous time translation symmetry (energy conservation law), and induces stationary amplification \cite{Yang_Hu_Li_Luo_2023,Lyubarov_Lumer_Dikopoltsev_Lustig_Sharabi_Segev_2022}, where the wave extracts energy from the external temporal modulation.

One of the interesting properties of PTCs is the superluminal group velocity modes on the band edge \cite{Sharabi_Lustig_Segev_2021,Pan_Cohen_Segev_2023,Martínez-Romero_Halevi_2017}. On the region adjacent to a momentum bandgap, the dispersion of PTCs is vertically flat. Thus, the band-edge mode could exhibit group velocity far beyond \textit{c\textsubscript{0}}. Remarkably, owing to the negligible loss of constituent materials, the pulse distortion is no longer an issue during the superluminal propagation. However, the band-edge modes in conventional PTCs have an extremely narrow bandwidth of momenta. As a result, the pulse can only propagate with a giant group velocity in the limit of plane-wave excitation at the band edge. Slightly enhancing the momentum bandwidth of the pulse will make a large portion of pulse momenta reside in the momentum bandgap, thereby inducing the parametric amplification to significantly slow down the group velocity of pulse. Up to date, how to realize the stable superluminal propagation of a finite-band pulse in linear PTCs still remains an open question. 

In the realm of static physical systems, band engineering stands as a potent tool for finely tuning the group velocity dispersion (GVD). Through spatial structure design, one gains the capability to manipulate the GVD at will, achieving intriguing phenomena such as flat bands, wherein the group velocity vanishes, and the eigenmodes exhibit localization properties across the entire Brillouin zone. Various approaches have been pursued to obtain flat band systems, including Lieb \cite{Vicencio_Cantillano_Morales-Inostroza_Real_Mejía-Cortés_Weimann_Szameit_Molina_2015} and Kagome \cite{Milićević_Montambaux__2019} photonic lattices in 2D, Moiré photonic superlattices in 1D \cite{Nguyen_Letartre_Drouard_Viktorovitch_Nguyen_Nguyen_2022} and 2D \cite{Tang_Ni_Du_Srikrishna_Mazur_2022,Mao_Shao_Luan_Wang_Ma_2021}, as well as three-dimensional flat Landau levels with pseudomagnetic fields \cite{cheng_three-dimensional_2024}, among others. These diverse methodologies hold promise for extension in space-time coordinates, thereby facilitating the engineering of Bloch-Floquet bands.

To this end, we propose the temporal photonic Moiré superlattice by transplanting the concept of photonic Moiré superlattice \cite{Nguyen_Letartre_Drouard_Viktorovitch_Nguyen_Nguyen_2022,Sunku_Ni_Jiang_Yoo_Sternbach_McLeod_Stauber_Xiong_Taniguchi_Watanabe_et_2018,Wang_Zheng_Chen_Huang_Kartashov_Torner_Konotop_Ye_2020} from space to time domain. As such, the flat energy band initially enabled by spatial counterpart is now transformed into the flat momentum band in the temporal photonic Moiré superlattice. Unlike flat energy band where the group velocity of flat-band modes is close to zero, the flat momentum band implies that the flat-band modes have an almost infinite group velocity in the entire Brillouin zone. As the momentum bandwidth of superluminal modes is favorably extended, the temporal photonic Moiré superlattice can enable the superluminal propagation of a finite-band pulse [Fig. 1(a)]. Furthermore, we prove that the superluminal pulse propagation in temporal photonic Moiré superlattice is highly stable with real-valued effective refractive index, as compared to active dispersive media where gain is used for compensation of resonance loss. Conceptually, such a flat momentum band is not restricted to the electromagnetic wave system but can be extended to any other common physical systems, such as water waves \cite{Bacot_Labousse_Eddi_Fink_Fort_2016,Apffel_Wildeman_Eddi_Fort_2022}, acoustic waves \cite{Shen_Li_Jia_Xie_Cummer_2019}, synthetic dimensions \cite{Long_Wang_Dutt_Fan_2023}, diffusion wave field \cite{Li_Zhang_Xu_Sun_Dai_Li_Qiu_2022}, etc.

\section{RESULT}

Inspired from the spatial counterpart, we construct the temporal photonic Moiré superlattice by superimposing two PTCs with different periods [Fig. 1(b)]. Without loss of generality, we consider two virtual photonic time crystals denoted as sublattice 1 \& 2. In each sublattice, the permittivity of two constituent materials is $\epsilon_1$ = 1  and $\epsilon_2$ = 3 and the filling ratio is 0.2. Moreover, the periods of sublattices 1\&2 are set as \textit{N} and \textit{N}+1, that satisfies $\tau$ = (\textit{N}+1)$\tau_1$ = \textit{N}$\tau_2$, where $\textit{N} \in \mathbb{Z}$. By applying the Boolean operator OR \cite{Harris_Harris_2010} to the high-index constituent material when we superimpose two sublattices, the temporal photonic Moiré superlattice is well-designed with the super period determined by $\tau$. 

\begin{figure}[htbp]
     \centering
     \includegraphics[width=1\linewidth]{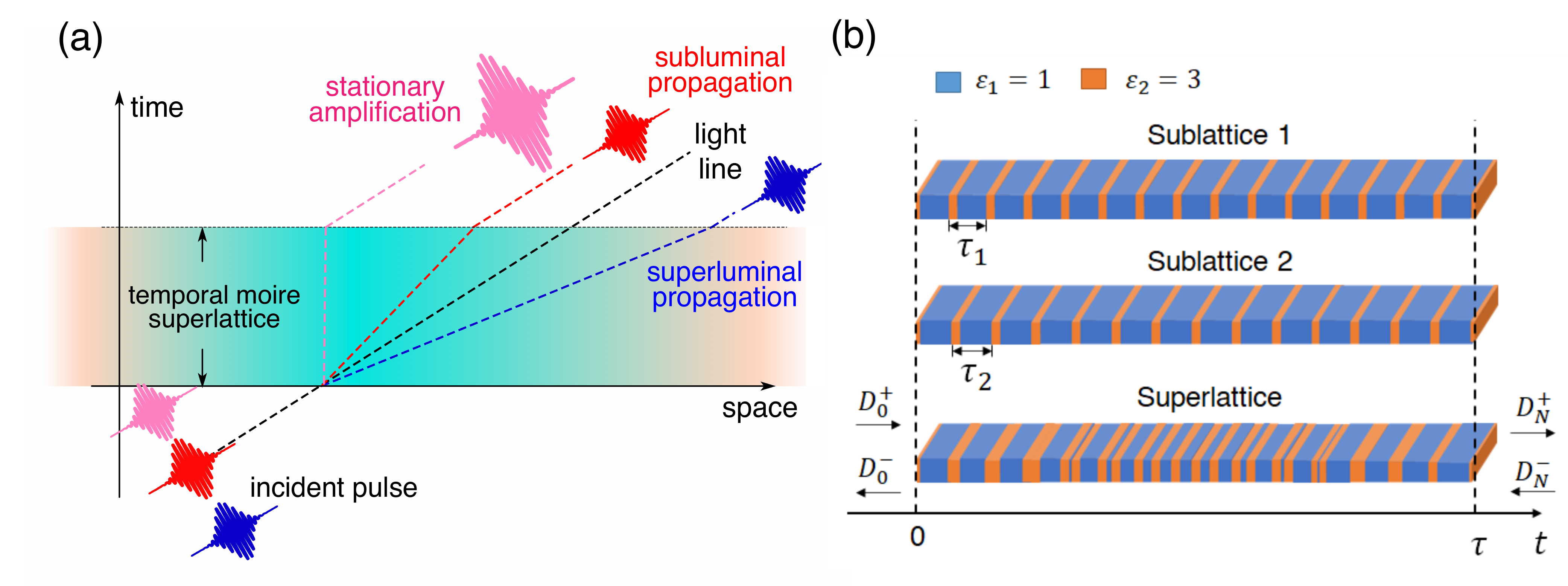}
     \caption{(a) Schematic of temporal photonic Moiré superlattice, for a pulse with different central momenta k, propagation inside the temporal photonic Moiré superlattice can either be superluminal, subluminal and stationary amplification. (b) Unit cell formation of temporal photonic Moiré superlattice by superposition of two sublattice with different period. $\tau$ = (\textit{N}+1)$\tau_1$ = \textit{N}$\tau_2$, $N$ = 15.}
    \end{figure} 
    
Next, we show that the temporal photonic Moiré superlattice can enable flat momentum band. To illustrate this, we plot the Floquet band structure of the temporal photonic Moiré superlattice in Fig. 2(a) (see supplementary section S1 for the calculation procedure), where two vertical flat bands emerge at 8.472$\textit{k}_0$ and 9.246$\textit{k}_0$, where $k_{0} = 2 \pi / \tau c_{0}$, analogous to the flattened electron-like band (bonding mode) and hole-like band (anti-bonding mode) in bilayer Moiré photonic superlattices \cite{Nguyen_Letartre_Drouard_Viktorovitch_Nguyen_Nguyen_2022} . In Spatial Photonic Crystal, flat bands support near zero group velocity and localized eigenmodes, where in each unit cell the intensity of electromagnetic field is confined around the localization center \cite{Wang_Zheng_Chen_Huang_Kartashov_Torner_Konotop_Ye_2020}. Similarly, on temporal flat bands of temporal photonic Moiré superlattice, localization properties emerge in the time domain. In Fig 2(b)-(d) we select three eigenmodes by fixing the momenta at normal band region, bonding flat band region and anti-bonding flat band region. In normal band region the eigenmode shows propagation profile where the energy is evenly distributed in space and time. Whereas on the two flat bands the eigenmodes is localized at either the center or two edges of the temporal unit cell. This temporal localization mode is fundamentally different from the localization mode in flat band Spatial Photonic Crystal. For periodic temporal modulation, the unit cell is set in time domain, that is, the energy of flatband eigenmode is temporally localized but evenly distributed in space. 
\begin{figure}[htbp]
     \centering
     \includegraphics[width=1.1\linewidth]{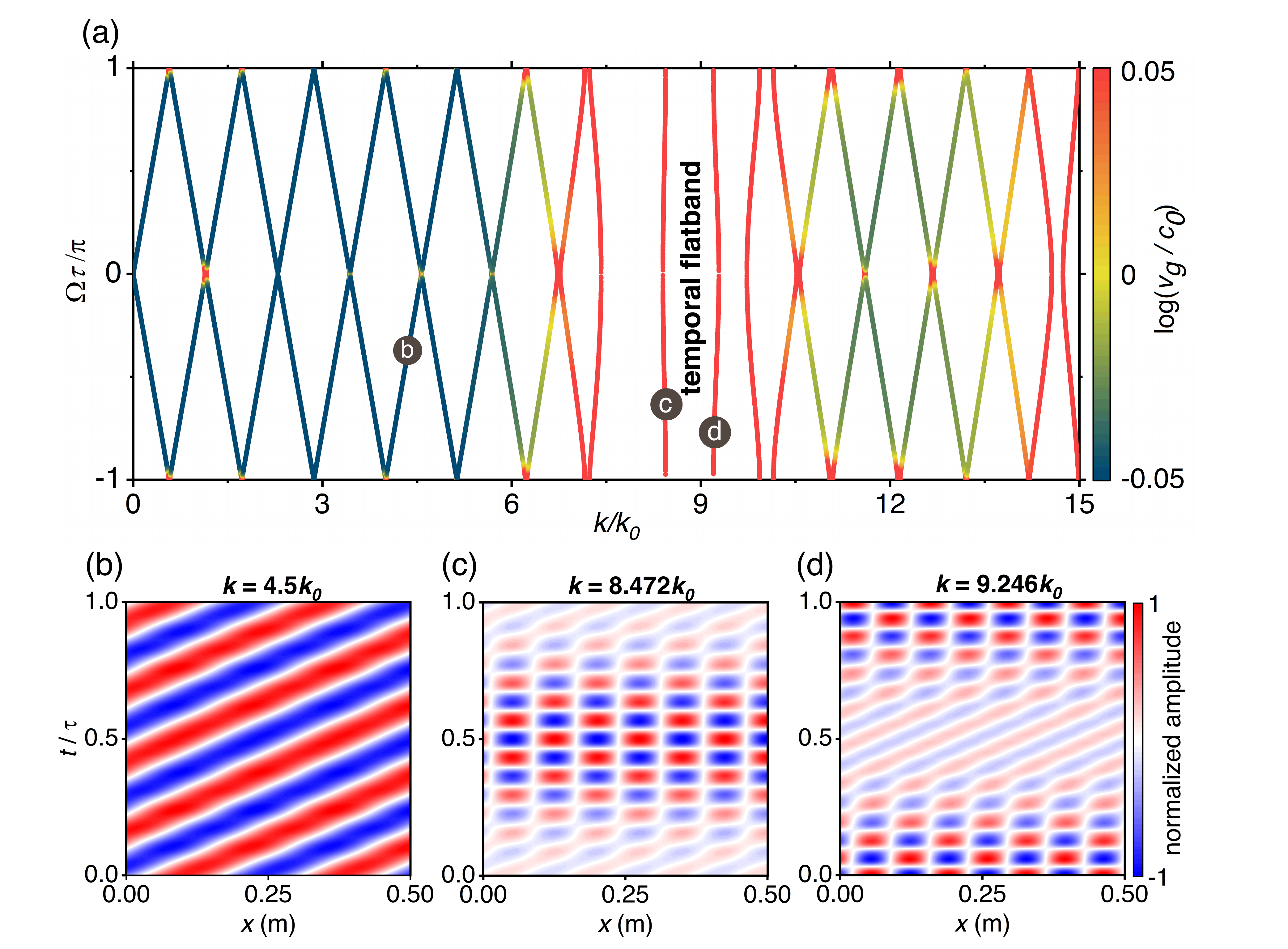}
     \caption{(a) Dispersion curves $\Omega(k)$ of temporal photonic Moiré superlattice. The dispersion curves are periodic along frequency axis hence only the 1st Brillouin zone is plotted. Two flat bands reside in k gaps, with group velocity $v_g$ indicated by color, which is superluminal at flat band region (red color) and subluminal at conventional band region (blue color). (b)-(d) Space-time eigenmode of temporal photonic Moiré superlattice at (b) normal band with $k=4.5k_{0}$ (c) flat band 1 with $k=8.472k_{0}$  (d) flat band 2 with $k=9.246k_{0}$, correspondingly. }
    \end{figure} 

Such a flat momentum band leads to the superluminal group velocity of flat-band modes. To demonstrate this, in Fig 2.(a) we denote the modes’ group velocity $v_{g} = \textrm{ } \partial \Omega / \partial k$ by the color of curves, demonstrating superluminal regions on the temporal flat band (red), and subluminal regions on normal bands (blue). As expected, the superluminal group velocity mode spans the band edge and the whole flat bands for their vertical flat dispersion. For the normal photonic bands, on the contrary, the group velocity is close to the effective phase velocity of the system. To study the dynamic evolution of different eigenmode in temporal photonic Moiré superlattice, we utilize Gaussian pulses as an input seed into our temporal photonic Moiré superlattice. In Fig.3(a)-(c), we present the numerically calculation result of three excitation dynamics base on Plane Wave Expansion (PWE) method. The momentum bandwidth of the pulse is equally set as $FWHM = 0.075 k_{0}$ for all three pulses, and the central momentum component is located at the center of the normal band ($4.5 k_{0}$), the band edge ($1.135 k_{0}$), and the center of the flat band ($8.42k_{0}$), respectively. Initially, all three pulses propagate in free space. At $t=0$, the modulation starts and the pulses enter photonic time Moiré superlattice from a temporal interface. After 20 cycles of modulation, the media returns to its original state $\epsilon_0$. When the pulse momenta centralize in normal band (Fig. 3(a)), the pulse experiences subluminal propagation. When the pulse momentum center locates at the band edge (Fig. 3(b)), although the band edge modes support superluminal group velocity, the band gap modes quickly induce stationary amplification and momentum gap mode dominates the total field. As a result, the wave packet of the pulse exhibits stationary and exponentially growing with zero group velocity. When the momentum component of incident pulse covers the momentum range of the flat band with all the eigenmodes supporting superluminal group velocities (Fig. 3(c)), the pulse center moves beyond the light line in both directions, showing omnidirectional superluminal pulse propagation. During the propagation inside temporal photonic Moiré superlattice, the pulse experiences a periodically gain and loss process with its spatial width expanded and suppressed, which is the consequence of the excitation of temporally localized modes. After the pulse transmits temporal photonic Moiré superlattice, the pulse maintains its original shape without distortion. Moreover, we plot the transmitted pulse time advance of three scenarios with different modulation periods in Fig. 3(d). Compared with vacuum propagation, both the normal band and the band edge excitation hold negative advance time. Only flat band excitation shows positive advance time, and linear relation with modulation cycle number.

Recently, there has been growing interest in broadband superluminal propagation based on active media \cite{Tsakmakidis_Reshef_Almpanis_Zouros_Mohammadi_Saadat_Sohrabi_Fahimi-Kashani_Etezadi_Boyd_et_2019}, which utilities stationary optical gain to engineer the frequency dispersion and compensate the loss from conventional anomalous dispersion region. However, the system has been shown to suffer from stability issue with a maximum propagation length before it forms amplifying oscillation for a finite superluminal bandwidth \cite{Duggan_Moussa_Ra’di_Sounas_Alù_2022}. As a system where the energy is not conserved, we propose that our system based on temporal photonic Moiré superlattice can greatly enhance the stability of superluminal propagation. To make comparison, we calculate the effective momentum dispersion for a finite temporal photonic Moiré superlattice slab. Here we consider a temporal photonic Moiré superlattice with 10 modulation periods and the effective complex refractive index and complex impedance for a given momentum plane wave can be retrieved via complex transmission functions $T$ and reflection functions$R$ \cite{Pacheco-Peña_Engheta_2020a} (also see supplementary section S2 for the derivation):

\begin{subequations}
\begin{align*}
e^{i \frac{c_{0} k \Delta t}{n_{e f f}}} &= \frac{1}{2} \sqrt{\frac{T^{4} + R^{4} + 1 - 2 T^{2} R^{2}  - 2 T^{2} - 2 R^{2}}{2 T^{2}}} \nonumber \\
&\quad + \frac{T^{2} - R^{2} + 1}{2 T}
\end{align*}

\begin{equation*}
\eta_{e f f} = \eta_{0} \frac{T^{2} - 2 T R - R^{2} + 1}{\sqrt{\left(T^{2} - R^{2} + 1\right)^{2} - 4 T^{2}}}
\end{equation*}
\end{subequations}

In Fig. 4(a) we plot the real and imaginary part of complex refractive index dispersion in the vicinity of the flat band region, where the imaginary part drops to zero in the band while turns negative in the band gap, corresponding to the amplification modes. To compare with the temporal photonic Moiré superlattice system, we also consider a double Inverted Lorentzian resonance (d-ILR) model: $\epsilon_{r} k = 1 + A k_{1}^{2} / (k_{1}^{2} - k^{2} - i \gamma_{1} k ) + \textrm{ } B k_{2}^{2} / ( k_{2}^{2} - k^{2} - i \gamma_{2} k )$. Since the frequency is not a conserved quantity when scattering at temporal interface, we set the permittivity here as a function of momenta  instead of frequency. The group velocity for dispersive materials is given by $v_{g} = R e (\frac{\partial \omega}{\partial k}) = c_{0} (\frac{1}{n^{'}} - \frac{1}{n'^{2}} \frac{\partial n^{\prime}}{\partial k})$, where $n (k) = n^{'} + i n^{''}$, under the approximation of $\frac{n^{''}}{n^{'}} \rightarrow 0$ which is valid in our designed bandwidth. To make a comparison we use algorism optimization to make both models share the same group velocity dispersion in the designed bandwidth. In Fig.4(b) we plot the calculated d-ILR group velocity and reference temporal photonic Moiré superlattice group velocity and the two dispersion curves fit well in the restricted bandwidth. Then we set the d-ILR as a temporal slab, where the refractive index switches from 1 to complex dispersive d-ILR at $t=0$, and switch back to 1 at $t = 10 \tau$. After numerical calculation we show the space-time evolution of the same gaussian beam in d-ILR in Fig. 4(c). We observe superluminal propagation where the center of pulse moves beyond the light line. However, after the pulse propagates through the critical length, where the instability takes the play, the propagation becomes unstable, and the pulse is strongly distorted. From the temporal profile comparison of outgoing wave in Fig 4(d) we observe sharp leap of displacement filed amplitude at the stable-unstable transition time. Taking $log\left(D_{z}\right)$ in the inset figure indicates that such transition is a consequence of the stationary gain by the imaginary part of the complex refractive index, which gives rise to an exponential amplification in time. In contrast, we found that temporal photonic Moiré superlattice can realize much stabler propagation owing to the modulation, as the effective imaginary part of refractive index in negligible and gives little fluctuation after each cycle of temporal photonic Moiré superlattice. This guarantees a stabilized superluminal broadband platform.
\begin{figure}[htbp]
     \centering
     \includegraphics[width=1\linewidth]{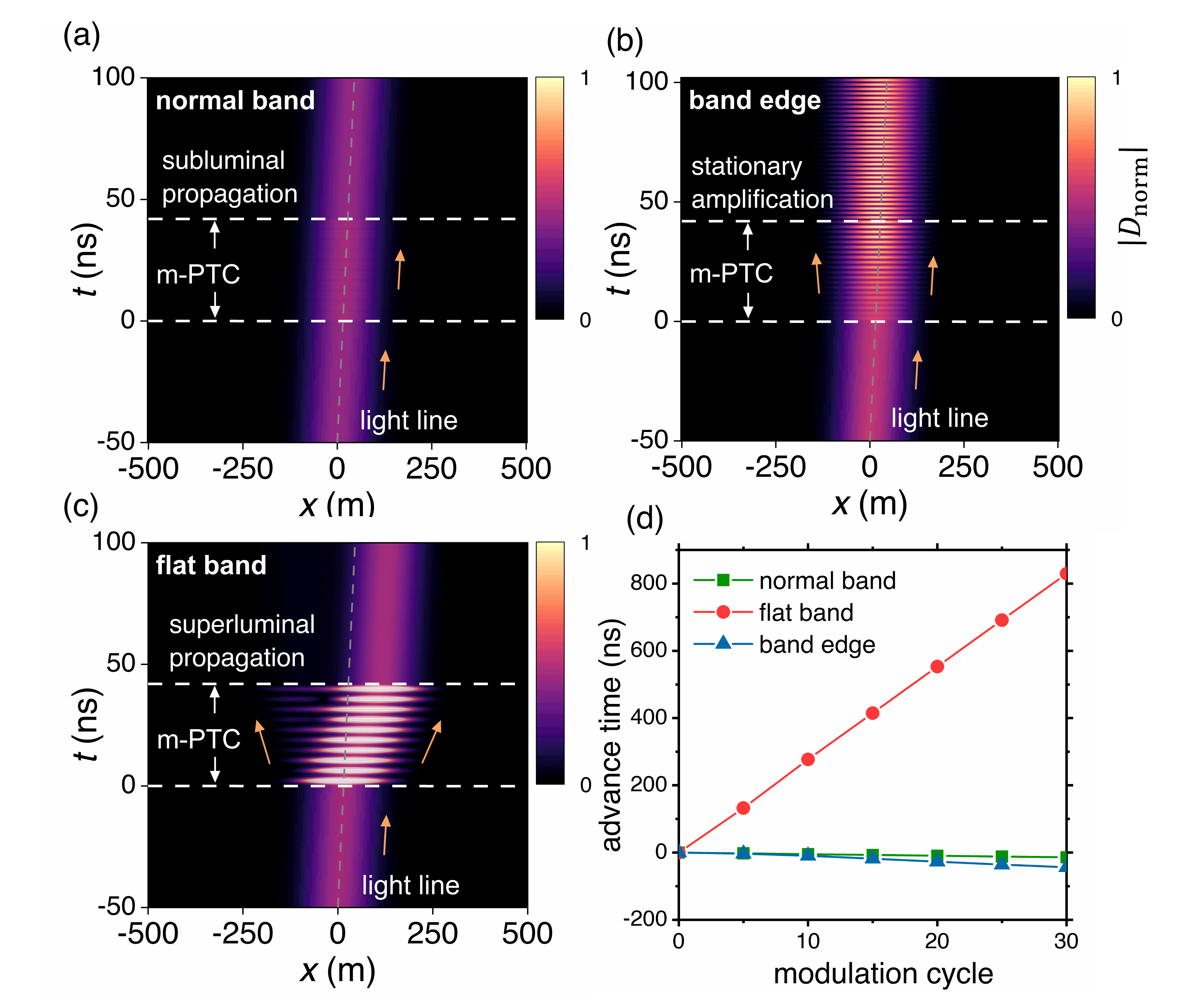}
     \caption{Numerically calculated Gaussian Pulse excitation in (a) normal band, (b) band edge and (c) flat band of temporal photonic Moiré superlattice (m-PTC), where the pulse exhibits subluminal propagation ($v_g=0.89c_0$), stationary amplification (peak does not move with wave front moving at $v=c_0$) and superluminal propagation ($v_g=8.08c_0$), respectively. (d) the outgoing pulse peak advance time under the modulation of different periods of Moiré photonic time crystal, compared with free space propagation without temporal modulation. Position selected at  x=700m.}
    \end{figure} 
\begin{figure}[htbp]
     \centering
     \includegraphics[width = 1.1\linewidth]{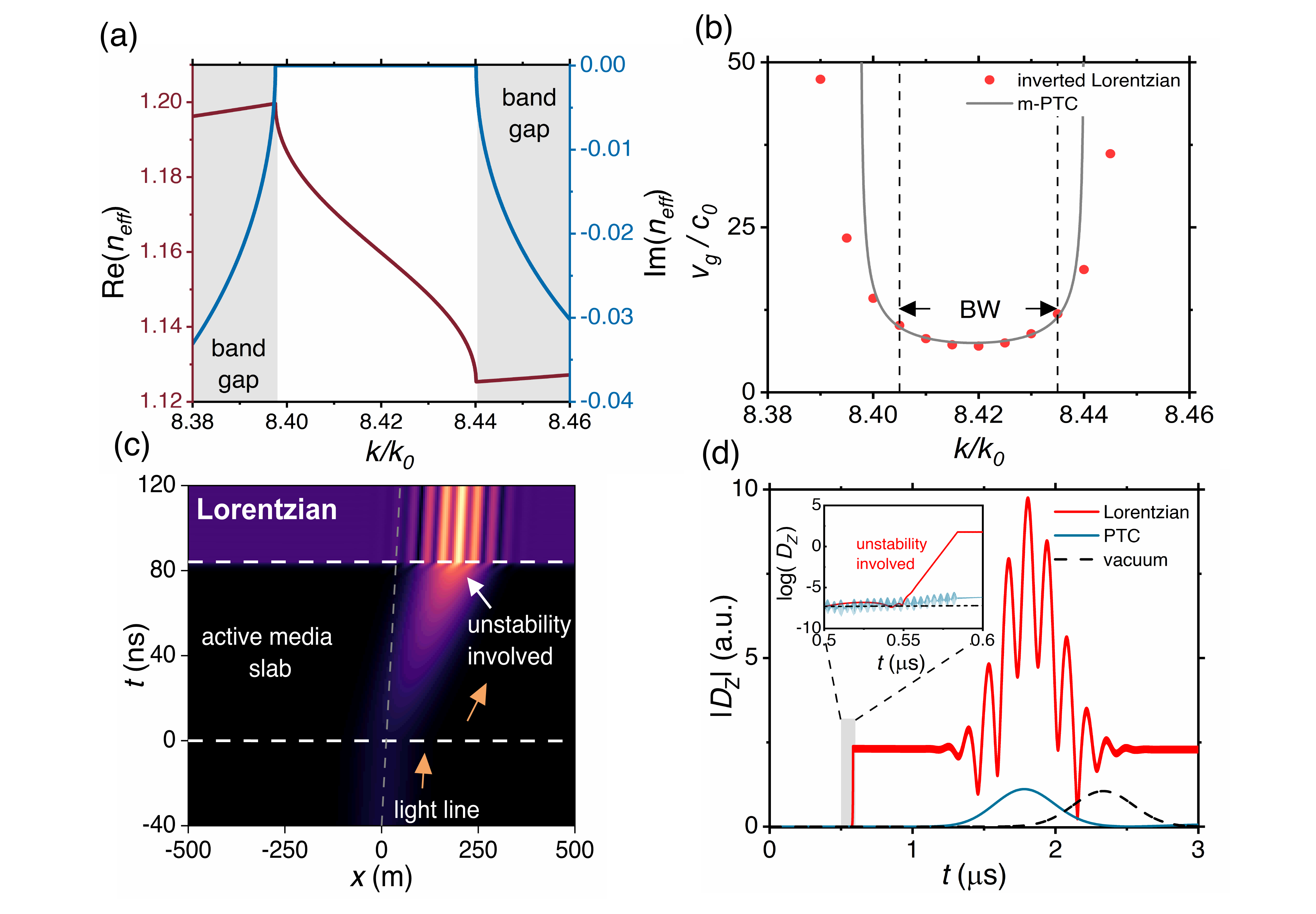}
     \caption{(a) Real part (solid) and imaginary part (dashed) of the retrieved effective refractive index. Inside the band region $Im(n_{eff})$ is close to zero, exhibit stable propagation. (b) group velocity dispersions for double inverted Lorentzian resonances and Moiré photonic crystal. (c) Space-time field distribution of double inverted Lorentzian resonances when a Gaussian pulse incident to an active slab with thickness of 10$\tau$, where the pulse undergoes unstable propagation and distortion. (d) The temporal profile of outgoing pulse after propagating through the identical temporal length of double inverted Lorentzian resonances slab and temporal photonic Moiré superlattice, the inset figure gives the zoomed field evolution during the modulation period.}
    \end{figure} 
The concept of forerunners in the context of superluminal propagation may raise concerns about the violation of Einstein's causality. However, in this letter, we utilized the plane wave expansion method, where a pulse is considered as a superposition of pre-existing plane waves in both space and time. Thus, there is no distinct forerunner for a given pulse. As demonstrated in \cite{Pan_Cohen_Segev_2023}, in temporally modulated system, any forerunner that exists will propagate at the speed of light in vacuum but will never exceed it. This is due to the momentum components of a sharp forerunner that encompass all the spatial wavevectors and the restriction that material response $\epsilon \left(\omega\right)$ tends towards unity as $\omega \rightarrow 0$ , For experimental validation of these phenomena, it is necessary to ensure that the seed pulse has a sufficiently large spatial profile to separate the forerunner from the center so the effect of causality can be suppressed. Several experimental platforms in microwave regime for investigating time varying system have been proposed, including transmission line, which can produce large switching induced temporal interface and temporal antireflection slabs \cite{Zhou_Alam_Karimi_Upham_Reshef_Liu_Willner_Boyd_2020} and temporal coherent control \cite{galiffi_broadband_2023}; Capacitive impenetrable metasurface \cite{Wang_Mirmoosa_Asadchy_Rockstuhl_Fan_Tretyakov_2023} and tunable Spoof surface plasmon polariton (SSPP) \cite{Zhang_Cui_Luo_Zhang_Xu_He_Zhang_2020,Gao_Zhang_Luo_Ma_Bai_Zhang_Cui_2021} can be utilized to study temporally modulation of surface wave.

\section{CONCLUSION}
In this letter, we have investigated the properties of flat bands in momentum space in the context of temporal photonic Moiré superlattice, revealing the nontrivial superluminal propagation of pulses in the flat band region, which has enhanced stabilization than conventional active media. which paves the way for extreme wave manipulation in more practical applications in highly efficient optical communication and computation. Note that we only modulated the media in time dimension, therefore our result also shows the possibility to extend it in space-time modulated system \cite{Sharabi_Lustig_Segev_2021,Galiffi_Huidobro_Pendry_2019,Huidobro_Galiffi_Guenneau_Craster_Pendry_2019} and anisotropic system \cite{He_Zhang_Qi_Bo_Li_2023,Li_Yin_He_Xu_Alù_Shapiro_2023}. Our work also sheds light on the potential of the study of PTCs, leading from binary structure \cite{Lustig_Sharabi_Segev_2018} to more complex superlattice unit cells designs and studying of PTC interacting with other systems \cite{Dikopoltsev_Sharabi_Lyubarov_Lumer_Tsesses_Lustig_Kaminer_Segev_2022}. 

\begin{acknowledgments}
This work is sponsored by National Research Foundation Singapore Competitive Research Program (NRF-CRP23-2019-0007).

\end{acknowledgments}


\bibliography{ref}

\end{document}